# Water/methanol solutions characterized by liquid µ-jet XPS and DFT – the methanol hydration case.

Authors

**Eric Pellegrin**[a†*], **Virginia Perez-Dieste**[a], **Carlos Escudero**[a], **Jordi Fraxedas**[a,b], **Pawel Rejmak**[c], **Nahikari Gonzalez**[a], **Abel Fontsere**[a], **Jordi Prat**[a] and **Salvador Ferrer**[a]

[a] CELLS-ALBA, Carrer de la Llum 2-26, Cerdanyola del Valles, Barcelona, 08290, Spain

[b] Catalan Institute of Nanoscience and Nanotechnology (ICN2), CSIC and BIST, Campus UAB, Bellaterra, 08193 Barcelona, Spain

[c] Institute of Physics PAS, Lotnikow 32/46, Warsaw, 02-668, Poland

Correspondence email: eric.pellegrin@zeiss.com

† Present address: Carl Zeiss SMT GmbH, Rudolf-Eber-Str. 2, Oberkochen, 73433, Germany

**Synopsis** The electronic structure of water and methanol as vapours, liquids, as well as liquid mixture is determined using synchrotron-based X-ray photoemission spectroscopy on liquid µ-jets. These results together with density functional theory provide, among others, interesting insight into the hydration of these two molecules in the liquid phase mixture.

**Abstract** The advent of liquid µ-jet setups as proposed by Faubel and Winter – in conjunction with X-ray Photoemission Spectroscopy (XPS) – has opened up a large variety of experimental possibilities in the field of atomic and molecular physics. In this study, we present first results from a synchrotron-based XPS core level and valence band electron spectroscopy study on water ($10^{-4}$M aqueous NaCl solution) as well as a water/methanol mixture using the newly commissioned ALBA liquid µ-jet setup. The experimental results are compared with simulations from density functional theory (DFT) regarding the electronic structure of single molecules, pure molecular clusters, and mixed clusters configurations as well as previous experimental studies. We give a detailed interpretation of the core level and valence band spectra for the vapour and liquid phases of both sample systems. The resulting overall picture gives insight into the water/methanol concentrations of the vapour and liquid phases as well as into the local electronic structure of the pertinent molecular clusters under consideration, with a special emphasis on methanol as the simplest amphiphilic molecule capable of creating hydrogen bonds.

**Keywords:** Liquid µ-jet XPS; water and aqueous alcohol solutions; synchrotron radiation spectroscopy; electronic structure.

# 1. Introduction

Starting from the study of water in its various phases, research within the field of atomic and molecular physics as well as especially the analysis of the electronic structure of liquids has gained a large momentum due to nowadays availability of sample environment techniques that allow for electron spectroscopy investigation using the previously existing set of experimental tools. Here, the seminal work performed by Siegbahn [1–4] on gases and liquids, together with that undertaken by Faubel & Winter [5–7] regarding the production of liquid jets and Salmerón & Schlögl [8] on near ambient pressure x-ray photoemission spectroscopy (NAP-XPS), respectively, have been ground breaking from the experimental point of view.

On the sample physics side, a central issue within the study of water and alcohols as well as their mixtures is the role of hydrogen bonds within the phase diagram of these systems, especially regarding their influence on structural properties of both liquids and solids on different intermolecular length scales. In fact, the actual structure of liquid water is still a matter of debate. Some authors have challenged the dominance of the well-stablished tetrahedral ordering in favour of asymmetrical structures with fewer than four hydrogen bonds per molecule based on X-ray absorption spectroscopies and theoretical calculations. [9] For a detailed analysis of photoemission (PES) results on pure water in its liquid and vapour phase we refer to the complete work by Faubel, Winter, and collaborators [7,10,11] as well as to the work by Fransson et al.[12] regarding a more general electron spectroscopy study on water. This also applies to the case of pure liquid/gaseous methanol and other alcohols investigated by Faubel.[5].

In this contribution, we present a combined experimental and theoretical study of the hydration of methanol at the aqueous solution/vapour interface with first results from the new liquid μ-jet setup at the Near Ambient Pressure Photoemission (NAPP) end station of the CIRCE beamline at the CELLS-ALBA synchrotron light source. Methanol is the simplest amphiphilic molecule capable of hydrogen bonding due to its (apolar) methyl and (polar) hydroxyl groups. In fact, methanol can be viewed as the result of substituting a hydrogen atom in a water molecule by a methyl group. This characteristic is one of the basic ingredients for the interest in the local structure of pure methanol in its liquid form, which naturally expands into the realm of water/methanol liquid mixtures. Starting from the methanol dimer model by Tomoda, [13] several models have been proposed for the molecular arrangements of these two molecular entities close to the liquid/vapour interface. [14–16] The interest in the molecular distribution of small polar molecules at aqueous solution/vapor interfaces has been recently extended to acetonitrile. [17]

Here, we make use of the salient features of synchrotron-based soft x-ray XPS such as, *e.g.*, small spot size, high brilliance, tuneable photon energy, etc. to enhance the O2s photoemission cross section with photon energies between 600 and 750 eV. XPS data from *pure* water (*i.e.*, $10^{-4}$ M NaCl aqueous solution) were also used as a necessary basis for the subsequent analysis of the XPS results from the methanol/water mixture.

## 2. Experimental

### 2.1. Near ambient pressure X-ray photoemission spectroscopy (NAP-XPS)

The experiment has been performed at the Near Ambient Pressure Photoemission (NAPP) end station of the CIRCE helical undulator beamline (100 – 2000 eV photon energy range) at the ALBA synchrotron light facility [18]. Both the photon beam entrance section and the entrance lens of the SPECS Phoibos 150 electron energy analyser at the NAPP experimental station are provided with differentially pumped stages that can establish a pressure ratio of $10^9$ between the sample and the electron analyzer as well as the upstream beamline optics. The electron analyser also includes extra pre-lenses to focus the photoelectrons through the small apertures between pumping stages. In order to make the orbital angular symmetry factor a constant, the XPS detection geometry fulfils the magic angle condition of 54.7º between the direction of polarization vector of the incoming monochromatic light and the electron analyser entrance lens axis.

The vertically deflecting toroidal refocusing mirror in the NAPP branch of the CIRCE beamline (located halfway between the monochromator exit slit and the sample position) includes a 1:1 entrance to exit arm configuration. Thus, the photon beam focal spot size at the μ-jet position is 19 (vert.) × 105 (hor.) μm$^2$ FWHM for a 20 μm exit slit setting (as used for the valence band XPS measurements – see below), which means that the μ-jet diameter of 30 μm roughly corresponds to the 4σ value of the vertical beam size (i.e., 32 μm).

For the XPS core level measurements, the beamline optics exit slit has been set to 50 μm and the analyser pass energy to 10 eV which results into a total electron energy resolution of 210 meV at 750 eV photon energy. This specific photon energy was chosen for (i) obtaining as high a photon flux possible for O1s XPS measurements as well as for (ii) increasing the O2p and O2s PES cross sections as compared to the corresponding C2p and C2s cross sections for the XPS valence band measurements (see ELETTRA home page for an overview on the photon energy dependence of the relevant photoemission cross sections[1]). For the latter valence band measurements, the monochromator exit slit size has been reduced to 20 μm with a resulting total electron energy resolution of 178 meV at 600 eV photon energy (i.e., resulting from a beamline photon and PES analyser electron energy resolution of 69.8 meV and 162.1 meV, respectively and including a thermal broadening of 25 meV). All photoemission measurements were performed at room temperature with the linear polarization vector of the incoming photon beam from the helical undulator within the horizontal plane that includes the µ-jet propagation direction. The reported binding energies have been determined within an error of ±0.07 eV.

---
[1] https://vuo.elettra.eu/services/elements/WebElements.html

The binding energy (BE) calibration of the XPS core level and valence band was performed using the distinct and well-established liquid water spectral features: 538.1 eV for the O1s line [11] and 11.16 eV for the $1b_1$ feature [7], respectively. Least-squares fits to the experimental XPS data were performed using the CasaXPS software (ww.casaxps.com) using a Shirley-type background.

**2.2. Liquid µ-jet system**

The liquid µ-jet setup used in this study consists of a commercial system manufactured by Microliquids GmbH (Göttingen, Germany) with the assistance by the CELLS Engineering Department (see Fig. 1). The mechanical layout is based on a "piggyback" design where the µ-jet positioning system is installed on top of the jet catcher manipulator. This allows for a non-critical "offline" relative jet/catcher alignment with the jet operating upfront the more critical approach to the XPS measurement position with a pre-aligned configuration. Both the jet and the catcher actuators are motorized and encoded along their pertinent three linear degrees of freedom.

The hydraulic part includes a HPLC pump-driven liquid system with quartz glass jet nozzles of different opening sizes and a jet catcher system evacuated by a downstream diaphragm pump. All fresh liquids were fed into the µ-jet system at room temperature, while the storage recipient for the liquid extracted by the µ-jet catcher was cooled to about 0º C and evacuated to about 1 mbar in order to keep the analysis chamber vacuum pressure low. In order to reduce the down time after an accidental icing of the jet catcher entrance opening, the latter is electrically heated in order to accelerate the melting process.

For the present study, the µ-jet operation parameters were as follows: 30 µm jet diameter, 1.5 ml/min. flow rate, 35 m/s jet flow speed, and 11 bar liquid pressure as provided by the HPLC pump. With these parameters, the typical stable jet length is 7 mm before decaying into a turbulent mode as well as a subsequent spray at larger distances from the jet nozzle.

Depending on experimental requirements and the selected vacuum pump configuration, the XPS analysis chamber vacuum pressures were set to 1.4, 1, and 0.003 mbar. The water sample consisted of degasified ultrapure Merck Milli-Q water with $10^{-4}$ M NaCl in order to reduce charging problems during the jet generation as well as photoemission process. The same type of ultrapure water was used for the preparation of the $\chi=0.21$ methanol/water solution (corresponding to 37.5% methanol).

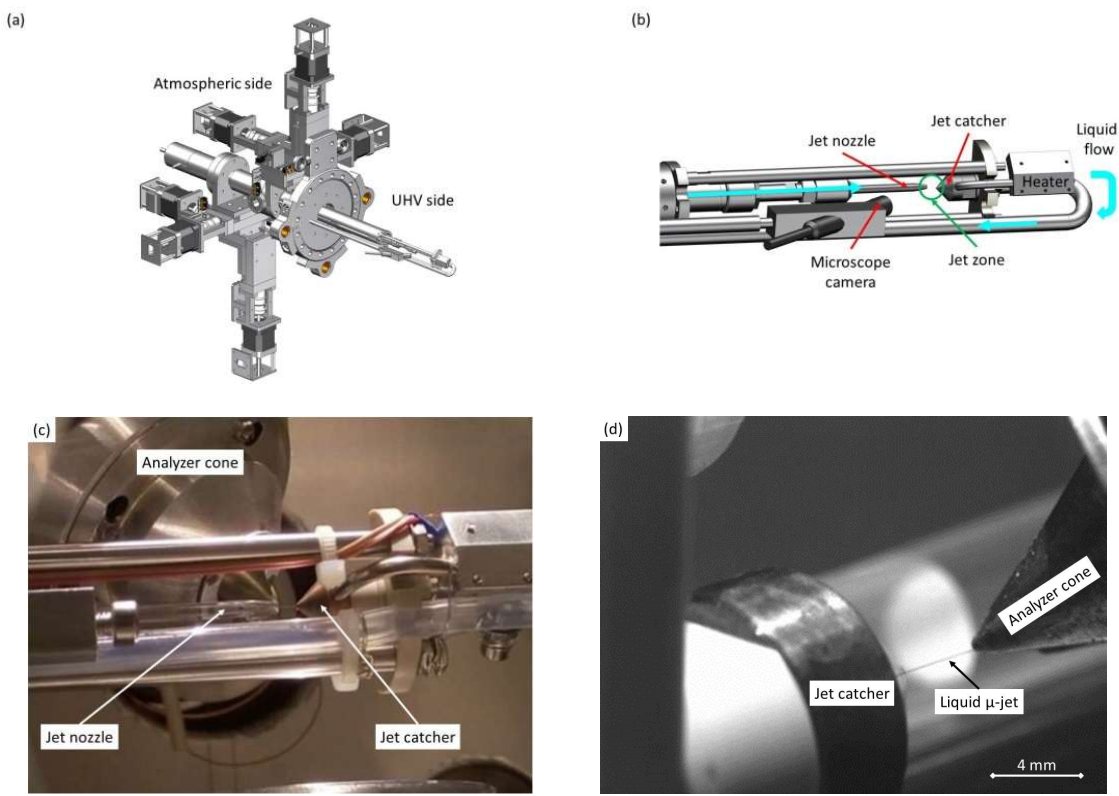

**Figure 1** (a, b) Technical drawings and (c, d) photographs of the liquid μ-jet setup at the CIRCE NAPP end station. (a) shows the in-air and in-UHV parts of the overall setup, whereas (b) shows a detailed view of the in-vacuum front part. In (c), the NAPP electron analyser entrance cone can be seen in the background on the left hand side while the jet nozzle and catcher can be seen in the centre of the image. In (d), the electron analyser entrance cone as well as the jet catcher can be seen on the right and left hand side, respectively. The liquid μ-jet with 30 μm diameter can be distinguished as a fine dark grey line between the analyser nozzle and catcher tips.

### 2.3. Density Functional Theory Calculations

The qualitative interpretation of XPS results has been supported by DFT calculations on simple cluster models (i.e., $(H_2O)_{1,3,5}$, $(CH_3OH)_{1,4}$, and $CH_3OH \cdot (H_2O)_3$, see Annex B1 and B2), performed with the Turbomole code.[19] A PBE0 hybrid functional [20] along with a def2-TZVP basis set [21] were applied in this study.

The BE can be calculated as the energy difference between the neutral and ionized molecule, or approximated as the negative orbital energies [22]. The former approach is formally strict; however, the convergence of the electronic structure calculation for the excited cation with hole below the top of the valence band is frequently difficult. In this work, core BEs were calculated using the first method, whereas the analysis of valence BEs is based upon Kohn-Sham orbital energies.

## 3. Results and discussion

### 3.1. XPS on liquid water

#### 3.1.1. Core level XPS

Figures 2(a) and 2(b) show the O1s XPS lines taken with 750 eV photons of a water μ-jet at an analysis chamber vacuum pressure of 1.4 and 0.003 mbar, respectively, with the monochromatic photon beam overlapping with the liquid μ-jet (*on-jet*). As in previous studies,[2,10,11,23] the liquid water contributes a broad peak at 538.1 eV BE [11] whereas the water gas phase contributes a narrow line around 540 eV BE (see Table 1).

As can be seen from these data, the area ratio between these two vapour/liquid XPS lines changes from 1.28 to 0.28 with decreasing chamber vacuum pressure. Here, with the water liquid at room temperature the vapour contribution shown for the 0.003 mbar spectrum together with the full jet/focused beam overlap is the minimum achievable, but could be reduced to the $10^{-5}$ mbar range when cooling down the incoming liquid water reservoir to 4º C and/or decreasing the jet diameter to 15 μm in view of decreasing the water vapour pressure [24].

The splitting between the two O1s lines for water molecules in both the liquid and the vapour phase is obtained as 1.6 and 1.9 eV at 0.003 and 1.4 mbar, respectively, and is in good agreement with a shift of 1.8 eV obtained earlier for liquid water μ-jets [11] and with theoretical and photoemission studies of free water clusters, with a reported shift of 1.6 eV when reducing the cluster size from 200 molecules (538.2 eV BE) to isolated molecules (539.8 eV BE). [25,26]

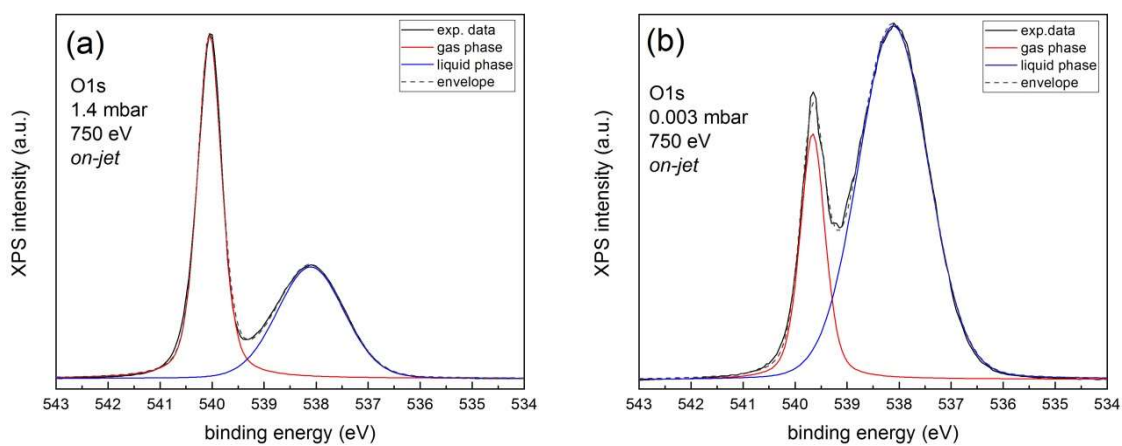

**Figure 2** O1s XPS lines of the water liquid μ-jet taken at 750 eV photon energy with the incoming monochromatic photon beam impinging onto the jet (*on-jet*) taken at a chamber pressure of 1.4 mbar (a) and 0.003 mbar (b), respectively. Continuous blue (liquid) and red (vapour) lines correspond to least-squares fits to the experimental data (continuous black lines) after a Shirley-type background subtraction using a weighted sum of Gaussian and Lorentzian functions. The dark-grey dashed line

corresponds to the envelope of the fit. The resulting parameters of the fit, BE, full width at half maximum (FWHM), hydration shift and vapour/liquid signal ratio, are given in Table 1.

**Table 1** Experimental binding energy (BE), FWHM, and gas-liquid shift figures obtained from fits to the H$_2$O experimental O1s XPS spectra shown in Fig. 2.

| Vacuum pressure | Vapour | | Liquid | | Hydration shift | Vapour/liquid ratio |
| --- | --- | --- | --- | --- | --- | --- |
| | BE | FWHM | BE | FWHM | | |
| 0.003 mbar | 539.7 | 0.57 | **538.1** | 1.59 | 1.6 | 0.28 |
| 1.4 mbar | 540.0 | 0.55 | **538.1** | 1.50 | 1.9 | 1.28 |

Bold numbers indicate BEs used for energy calibration. BE, FWHM, and shift values are given in eV.

Interestingly, a significant BE shift can be observed between the vapour peaks at the different vacuum pressures; BE increases by 0.3 eV as pressure increases from 0.003 mbar to 1.4 mbar (see Table 1). The same shift is observed for the VB gas features (not shown). From the raw data we observe that the kinetic energies for the liquid phase do not change with pressure, which should, in principle, rule out any contributions from changes in the surface potential [27,28]. We hypothesize that the observed shift arises, at least in part, from a change of the chemical potential ($\mu$) of the gas which would lead to a change in the binding and kinetic energies of the photoelectrons. A simple estimation based on the ideal gas approximation gives an energy shift of about 0.15 eV at room temperature using the expression $\Delta\mu = k_B T \ln(p_1/p_2)$, where $k_B$ is the Boltzmann constant, $T$ the temperature and $p_1$=1.4 mbar and $p_2$=0.003 mbar, respectively. This simple approach accounts for half of the observed 0.3 eV shift. More realistic gas models do not provide any significant contribution to the estimated value. To this point, we cannot be more precise on the origin of the observed 0.3 eV shift and future dedicated experiments should be undertaken.

### 3.1.2. Theoretical O1s binding energy analysis

The DFT simulations predict a decrease of the O1s BE for water clusters with respect to the gas phase (i.e., isolated single water molecules), as previously discussed in the literature, [26] although the BE shift value is somewhat underestimated (see Table 2). The decrease in O1s BE correlates with the increase of O-H bond lengths and H-O-H bond angle in hydrogen bound water molecules.

**Table 2**  DFT computed O1s binding energies and binding energy shifts (BE and ΔBE, in eV), selected bond distances (R, in Å) and bond angles (θ, in degrees).

| | $BE_{O1s}/\Delta BE$ (eV) | $R_{OH}$ (Å) | $R_{H \rightarrow O}$ (Å) | $R_{O \leftarrow H}$ (Å) | $\theta_{HOH/HOC}$ (°) | $R_{CO}$ (Å) |
|---|---|---|---|---|---|---|
| $H_2O_{gas}$ | 539.3 | 0.960 | - | - | 105.07 | - |
| $(H_2O)_3$ | 538.7/−0.6 | 0.959, 0.977[1] | 1.865 | | 106.15 | - |
| $(H_2O)_5$[2] | 538.3/−1.0 | 0.976 | 1.872 | 1.890 | 107.28 | - |
| $CH_3OH_{gas}$ | 538.6 | 0.959 | - | - | 108.51 | 1.409 |
| $(CH_3OH)_4$ | 537.8/-0.8 | 0.986 | 1.715 | | 109.28 | 1.410 |
| $CH_3OH(H_2O)_3$ | -[3] | 0.984 | 1.733 | 1.710 | 108.97 | 1.411 |

[1]OH bond for the H atom involved in hydrogen bond to the next $H_2O$ molecule. [2]All data for the central, tetrahedrally coordinated $H_2O$ molecule in $(H_2O)_5$ cluster. [3]State with core-hole at O atom in $CH_3OH$ not converged.

### 3.1.3. Valence band XPS

Figures 3(a) and 3(b) show the XPS valence band (VB) spectra for the water vapour and liquid phase, respectively, taken at 600 eV photon energy with an electron analyser pass energy of 10 eV and with a 20 μm exit slit. The VB spectrum from Fig. 3(b) was obtained from a weighted subtraction of the gas phase spectrum from the measured total raw VB data (see Fig. 8 in Annex A1). As has been discussed earlier, [29] a BE shift as well as a broadening of gas phase XPS lines within mixed gas-liquid phase spectra can be observed. In the present case, a consequence from this can be observed in the VB spectrum of liquid water shown in Fig. 3(b), where the subtraction of the *off-jet* water gas phase VB spectrum from the *on-jet* mixed vapour/liquid spectrum results in a non-zero residue of the $1b_1$ gas line within the liquid VB spectrum around 13 eV BE. This is very probably due to the aforementioned broadening of the $1b_1$ gas line in the *on-jet* gas spectrum as observed earlier.

The gas phase data exhibit the well-known $1b_1$, $3a_1$, and $1b_2$ lines, at 12.82, 15.0 and 18.8 eV, respectively, as shown in Table 3, plus two additional resolved lines on the high BE side of the $1b_1$ line at 13.2 and 13.6 eV BE, which we attribute to the first and second inelastic scattering, respectively, from the symmetric/asymmetric vibrational stretching modes of the isolated water molecules. The stretch shifts shown in Table 3 correspond fairly well to experimental values from vibrational analysis (symmetric = 0.453 eV, asymmetric = 0.466 eV; the bending mode accounts only for 0.198 eV). [30]

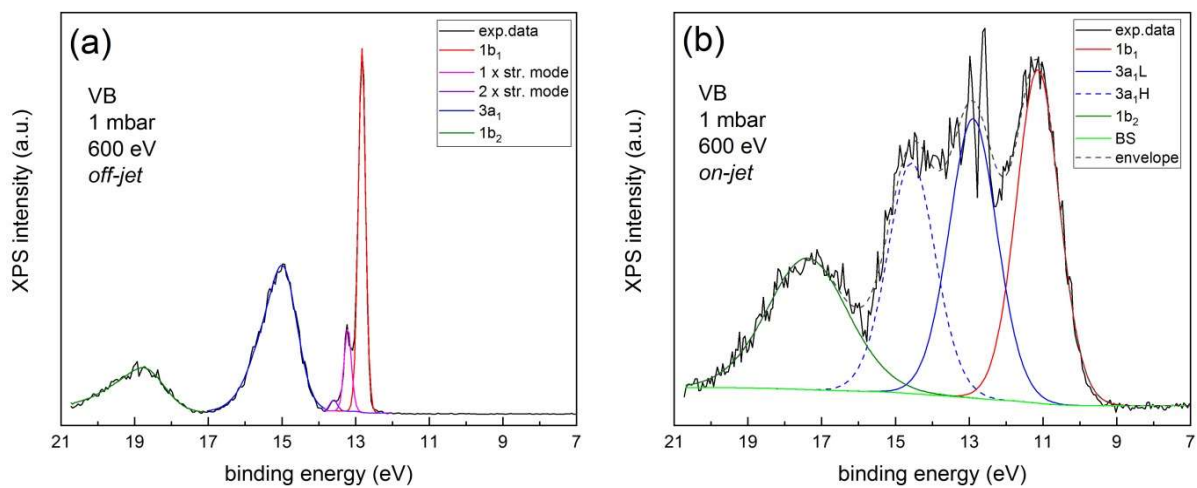

**Figure 3** Valence band XPS data of the water liquid µ-jet taken at 600 eV photon energy with a chamber vacuum pressure of 1 mbar. The data were taken with the photon beam passing (a) above the liquid µ-jet (*off-jet*) and (b) with the incoming monochromatic photon beam impinging onto the jet (*on-jet*) and after subtraction of a weighted gas phase spectrum. Least-squares fit to the experimental data (continuous black line) after a Shirley-type background subtraction are given. In (a) the $1b_1$ and $3a_1$-$1b_2$ regions are deconvoluted separately using symmetric and asymmetric functions, respectively, an asymmetry that accounts for unresolved inelastic scattering. In (b) four symmetric components are found, which are associated to the $1b_1$, $3a_1$ and $1b_2$ molecular orbital character. The resulting parameters from the fit are given in Table 3.

The liquid water VB spectrum shown in Fig. 3(b) compares well with previously published data taken at photon energies between 516 and 600 eV, where the O2s photoemission cross section is well beyond that of O2p. Taking a closer look at the numbers in Table 3, we would like to point out the good overall correspondence with the study by Winter et al. [7] taken at 60 eV photon energy. Also, making use of the above enhanced O2s photoemission cross section at 600 eV photon energy for specifically the $3a_1$ MO – which includes a significant O2s orbital character (see Annex B1) – we could derive a $3a_1$L-$3a_1$H splitting of 1.67 eV, which can be compared to the values of 1.55 eV and 1.38 eV by Nordlund et al. for $H_2O$ ice and Nishizawa et al. for $H_2O$ liquid, respectively [23,30]. Regarding the physical interpretation of this $3a_1$L-$3a_1$H splitting, we would like to put forward that it may essentially arise from the short-range tetrahedral distribution of the water molecules, which is why it is observed in both ordered (i.e., crystalline ice) and disordered systems (i.e., amorphous ice).[32,33] The splitting in the $3a_1$ band can be further clarified via DFT results as arising from the covalent component of intermolecular hydrogen bonds: here, $3a_1$L and $3a_1$H are due to $3a_1$ orbitals of neighboring $H_2O$ molecules overlapping in an antibonding and bonding like manner (see the pertinent $3a_1$ MOs in Fig. 11 of Annex B1; center panel, middle and lower row). Similar effects are found in DFT models for the $1b_2$ band, which can contribute to the observed broadening of this band in liquid water. [31]

In Table 3, we present the gas-liquid BE shifts derived from the fits to the data shown in Fig. 3 as compared to equivalent figures measured at 60 eV photon energy. [7] From this comparison, one can see that there is a good overall correspondence regarding these experimental shifts. This also applies to the line widths when going from the vapour to the liquid phase, reflecting the order/disorder phenomena in the latter.

**Table 3** Experimental binding energy, FWHM, and gas-liquid shift figures obtained from the fits to $H_2O$ experimental valence band spectra shown in Fig. 3.

| Orbital | Vapour | | | Liquid | | | Gas-liq. shift | Gas-liq. shift[2] |
|---|---|---|---|---|---|---|---|---|
| | BE | Stretch shift | FWHM | Ref. BE[3] | BE | FWHM | | |
| $1b_1$ | 12.82 | | 0.23 | 12.62 | **11.16** | 1.38 | 1.66 | 1.45 |
| Stretch 1st | 13.20 | 0.38 | 0.23 | - - | - - | - - | - - | |
| Stretch 2nd | 13.59 | 0.77 | 0.23 | - - | - - | - - | - - | |
| $3a_1/3a_1L$ | 15.0 | | 0.95 | 14.78 | 12.90 | 1.53 | 1.26[1] | 1.34 |
| $3a_1H$ | - - | | - - | - - | 14.57 | 1.53 | | |
| $1b_2$ | 18.8 | | 1.32 | 18.55 | 17.40 | 2.57 | 1.4 | 1.46 |

[1]Averaged shift of 3a1gL and 3a1gH peaks.  [2]From [7]  [3]From [3]

Bold numbers indicate binding energies used for binding energy calibration. BE, FWHM, and shift values are given in eV.

## 3.2. XPS on methanol(37.5%) / water mixture

### 3.2.1. Core level XPS

Fig. 4 shows the O1s XPS lines for the methanol (37.5%)/water mixture measured at a chamber vacuum pressure of 1 mbar for the gas phase [Fig.4(a)], as measured with the photon beam passing beyond the liquid μ-jet (*off-jet*), as well as for the *on-jet* case where the photon beam overlaps with the liquid μ-jet [Fig. 4(b)]. As in the case of the *pure* water jet, the latter XPS spectrum shows the contributions from both the vapour and the liquid phases. Both spectra are in good qualitative agreement with the early data on the same sample system by Starr *et al.* [34]

Taking a closer look at the gas phase spectra in Fig. 4(a), one can clearly distinguish the O1s lines of O in water vapour and methanol vapour at 540.0 eV and 539.16 eV, respectively, which yields an O1s binding energy shift of 0.84 eV between these two molecules and which is in good agreement with previous findings of 0.85 eV. [34] Note that this is a considerable O1s BE shift induced by replacing a hydrogen atom within a water molecule by a methyl group. Furthermore, the broad line associated to O in the liquid methanol/water mixture can be observed in the *on-jet* spectrum at 538.0 eV BE in Fig. 4(b) with a FWHM of 1.38 eV (see Table 4). Interestingly, the line width differs from the corresponding value found for liquid water (*i.e.*, 538.1 eV BE, 1.50 eV FWHM; see Table 1).

The reduction in O1s XPS line width by roughly 0.2 eV when going from liquid water to the liquid methanol/water mixture is somewhat surprising. Instead, one would rather expect a line broadening due to the fact that the latter sample system includes two chemically inequivalent oxygen species. This can possibly be explained by short range ordering phenomena (*e.g.*, hydration) present in the liquid methanol/water mixture not being present in liquid water. We speculate that the amphiphilic character of the methanol molecule should play a relevant role here since the surrounding water molecules would obviously tend to more closely interact with the polar methanol hydroxyl group (*i.e.*, less than with the apolar methyl group), thus resulting into the short range ordering phenomena mentioned above.

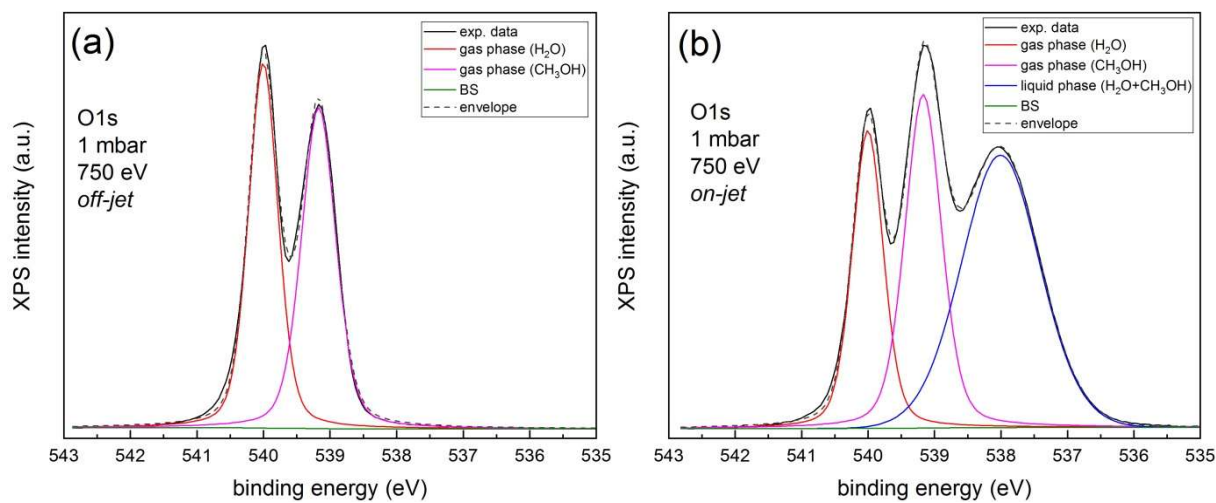

**Figure 4** O1s XPS lines of the methanol (37.5%)/water liquid μ-jet taken at 750 eV photon energy at a chamber pressure of 1 mbar. The data were taken with the photon beam (a) passing above the liquid jet *(off-jet)* and (b) impinging onto the jet *(on-jet)*. Least-squares fits to the experimental data (continuous black line) after a Shirley-type background subtraction are given. The water and methanol gas phase contributions are represented by continuous red and magenta lines, respectively, while the liquid mixture by a continuous blue line. The resulting parameters from the fit are given in Table 4.

**Table 4** Experimental binding energy, FWHM, and gas-liquid shift figures obtained from fits to the methanol/water experimental O1s XPS spectra shown in Fig. 4.

| Beam-jet interaction | Water vapour | | Methanol vapour | | BE shift | Liquid mixture | | $CH_3OH/H_2O$ vapour ratio |
|---|---|---|---|---|---|---|---|---|
| | BE | FWHM | BE | FWHM | | BE | FWHM | |
| *off-jet* | 540.0 | 0.52 | 539.16 | 0.61 | 0.84 | - - | - - | 1.02 |
| *on-jet* | 540.0 | 0.55 | 539.16 | 0.65 | 0.84 | 538.0 | 1.38 | 1.32 |

BE, FWHM, and shift values are given in eV.

In Fig. 5, we report on the C1s XPS lines on the methanol (37.5%)/water mixture measured under the same experimental conditions as the O1s data in Fig. 4 (*i.e.*, *off-jet* and *on-jet* for gas phase and gas/liquid phase, respectively). Both spectra show the narrow C1s XPS line associated to the methanol C atom present in the vapour phase with a BE of about 292.5 eV, including a series of vibrational lines on the high BE side of the main C1s line [*i.e.*, for n = 0, 1, 2, 3; see Fig. 5(a)] with energy losses of about 0.4 eV due to the C-H stretching mode of the methyl group. This value is in good agreement with energies of 0.39 eV, as determined by XPS for methanol on Cu(100) [35] and 0.36 eV from infrared measurements on methanol vapour [36]. The broad C1s line at 291.56 eV BE in Fig. 5(b) is obviously associated to liquid methanol, thus yielding a methanol gas-liquid C1s BE shift about 0.95 eV (see Table 5).

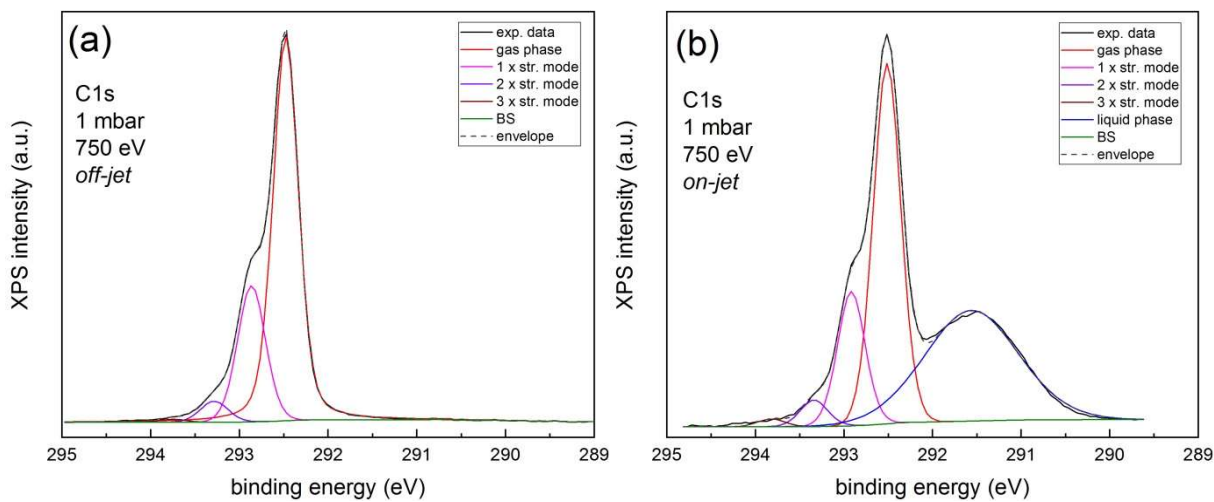

**Figure 5** C1s XPS lines of the methanol (37.5%)/water liquid μ-jet taken at 750 eV photon energy at a chamber pressure of 1 mbar. The data were taken with the photon beam (a) passing above the liquid jet (*off-jet*) and (b) impinging onto the jet (*on-jet*). Least-squares fits to the experimental data (continuous black line) after a Shirley-type background subtraction are given. The methanol vapour and liquid phases are represented by continuous red and blue lines, respectively. The resulting parameters from the fit are given in Table 5.

From a quantitative analysis of the peak area ratios in the above O1s core level spectra, we can now derive the gas phase methanol volume concentration for both the *off-jet* and the *on-jet* experimental configuration. Starting from the methanol/water O1s line ratios of 1.02 and 1.32 for the *off-jet* and *on-jet* configuration (see Table 4), respectively, we obtain the corresponding figures for the methanol volume concentrations of 82.5% (equivalent to a methanol mole fraction of $\chi = 0.50$) and 85.9% (equivalent to $\chi = 0.57$). The increase in $\chi$ when going from the *off-jet* and *on-jet* configuration can be readily understood based on an increasing methanol concentration gradient with the analyser entrance lens axis aiming at the μ-jet core in the latter *on-jet* configuration. The *on-jet* $\chi$ number is also in qualitative agreement with the findings by Starr *et al*. [34] using a room temperature droplet train setup at 2.5 Torr (3.33 mbar) vacuum pressure yielding $\chi = 0.56$ in the vapour phase and corresponding to a $\chi_{vapour} = 0.62$ at 35º C for a $\chi_{liquid} = 0.21$ methanol/water liquid mixture as used in our experiments.[37] A similar quantitative analysis of the XPS valence band spectra given in the next section will basically confirm the above results for the methanol/water vapour phase.

**Table 5** Experimental binding energy, FWHM, and gas-liquid shift figures obtained from fits to the methanol/water experimental C1s XPS spectra shown in Fig. 5 (all numbers are given in eV).

| *off-jet* | BE | vibr. shift | FWHM | *on-jet* | BE | vibr. shift | FWHM | Gas-liq. shift |
|---|---|---|---|---|---|---|---|---|
| n = 0 (vapour) | 292.48 | - - | 0.34 | n = 0 (vapour) | 292.51 | - - | 0.37 | 0.95 |
| n = 1 (vapour) | 292.86 | 0.38 | 0.37 | n = 1 (vapour) | 292.92 | 0.41 | 0.36 | - - |
| n = 2 (vapour) | 293.28 | 0.8 | 0.37 | n = 2 (vapour) | 293.33 | 0.82 | 0.36 | - - |
| n = 3 (vapour) | 293.75 | 1.27 | 0.37 | n = 3 (vapour) | 293.80 | 1.29 | 0.36 | - - |
| liquid | - - | - - | - - | liquid | 291.56 | - - | 1.24 | - - |

BE, FWHM, and shift values are given in eV.

### 3.2.2. Valence band XPS

In Fig. 6, we present the XPS valence band spectra of methanol vapour [Fig. 6(a)] and methanol/water liquid [Fig. 6(b)] taken at 600 eV photon energy (10 eV electron analyser pass energy) and at a chamber vacuum pressure of 1 mbar. The vapour VB spectrum in Fig. 6(a) has been obtained by a weighted subtraction of the water vapour VB data [see Fig. 3(a)] from the total *off-jet* VB raw data (see Annex A2, Fig. 9). In the same vein, the XPS VB spectrum of methanol in the methanol/water mixture in Fig. 6(b) was obtained from the total *on-jet* VB raw data (*i.e.*, with the incoming monochromatic photon beam overlapping with the µ-jet). The VB spectrum of liquid methanol in water shown was obtained after subtraction of the weighted methanol vapour spectrum [Fig. 6(a)] and the weighted water gas and liquid VB spectra (see Fig. 3 and Appendix A3, Fig. 10).

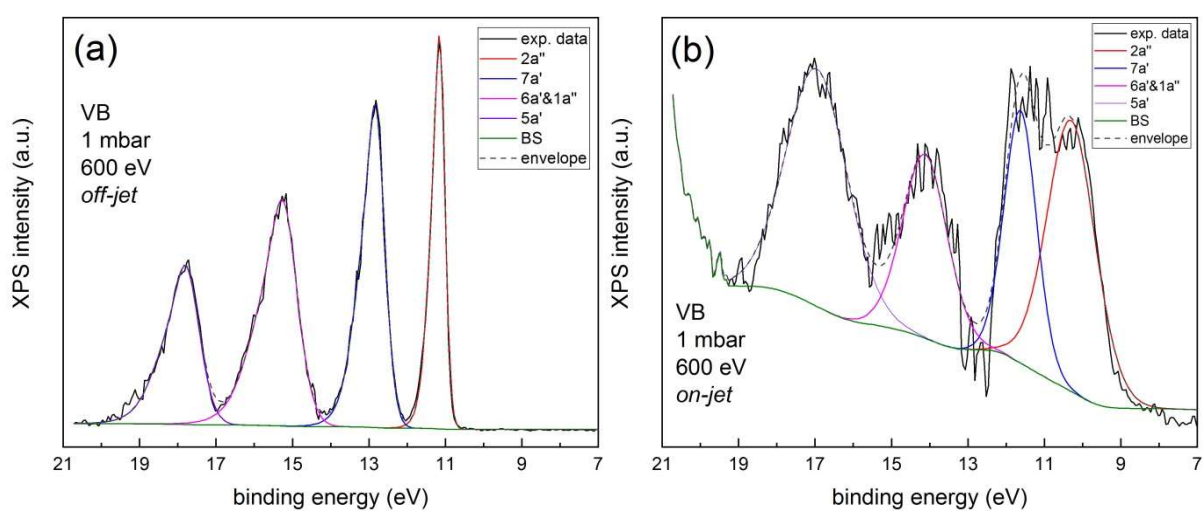

**Figure 6** Valence band XPS data from the methanol (37.5%)/water liquid µ-jet taken at 600 eV photon energy at a chamber pressure of 1 mbar. (a) Data taken with the photon beam passing above the liquid µ-jet (*off-jet*); the methanol vapour VB spectrum shown was obtained by subtraction of the water vapour valence band spectrum. (b) Data obtained with the incoming monochromatic photon beam overlapping with the µ-jet (*on-jet*); the VB spectrum of liquid methanol in water shown was obtained after subtraction of the weighted methanol vapour spectrum from Fig. 6(a) as well as of the water gas and liquid VB spectra. The least-squares fits to the experimental data (continuous black line) after a Shirley-type background subtraction are shown and the corresponding molecular orbitals identified: continuous red line (2a''), continuous blue line (7a'), continuous magenta line (6a' and 1a'') and continuous violet line (5a'). In (a) asymmetric Gaussian-Lorentzian functions have been used to cope with inelastic scattering while in (b) symmetric functions have been used. The resulting parameters from the fit are given in Table 6.

**Table 6** Experimental binding energy, FWHM, and gas-liquid shift figures obtained from the fits to methanol experimental valence band spectra shown in Fig. 6.

| Orbital | Methanol vapour | | Methanol / water liquid | | Shift gas-liq. mixture | Shift methanol gas-liq.[1] |
|---|---|---|---|---|---|---|
| | BE | FWHM | BE | FWHM | | |
| 2a" ($n_0$) | 11.16 | 0.34 | 10.30 | 1.49 | 0.86 | 0.95 |
| 7a' ($\underline{n}_0$) | 12.84 | 0.59 | 11.63 | 0.99 | 1.21 | 0.40 |
| 6a' & 1a" ($\sigma_{CO}$ & $\pi_{CH3}$) | 15.28 | 0.92 | 14.12 | 1.37 | 1.16 | 0.66 |
| 5a' ($\sigma_{OH}$) | 17.81 | 0.86 | 16.95 | 1.88 | 0.86 | 1.49 |

[1]From: [5]. BE, FWHM, and shift values are given in eV.

The methanol vapour VB spectrum in Fig. 6(a) exhibits four peaks of different increasing widths associated to the first five methanol molecular orbitals (see Table 6) and is compared with the corresponding ultraviolet photoemission spectroscopy (UPS) VB data by Faubel et al.[5] taken with 21.22 eV photons (He I) in Fig. 7 (a). We note that there is a good agreement between the BEs reported in previous studies.[5,13] While the overall spectral signatures are similar, the relative intensities of the lines are strongly altered as well as the symmetry of the peak at 15.28 eV BE decreases when going from 600 eV (continue green line) to HeI (full blue dots) incoming photon energy. This can be readily explained by the different C and O orbital characteristics of the methanol MOs involved in conjunction with the variation of photoemission cross sections for these two different photon energies. In some more detail, the former overall changes in relative peak intensities can be rationalized by the strong decrease of the C2p PES cross section together with a similar increase in O2s cross section for 600 eV photons. Also, the symmetry change of the peak at about 15.28 eV BE can be related to the increase of the calculated cross section intensity ratio between the 6a' and the 1a" MOs (see Figs. 13 and 14 in Annex B3).

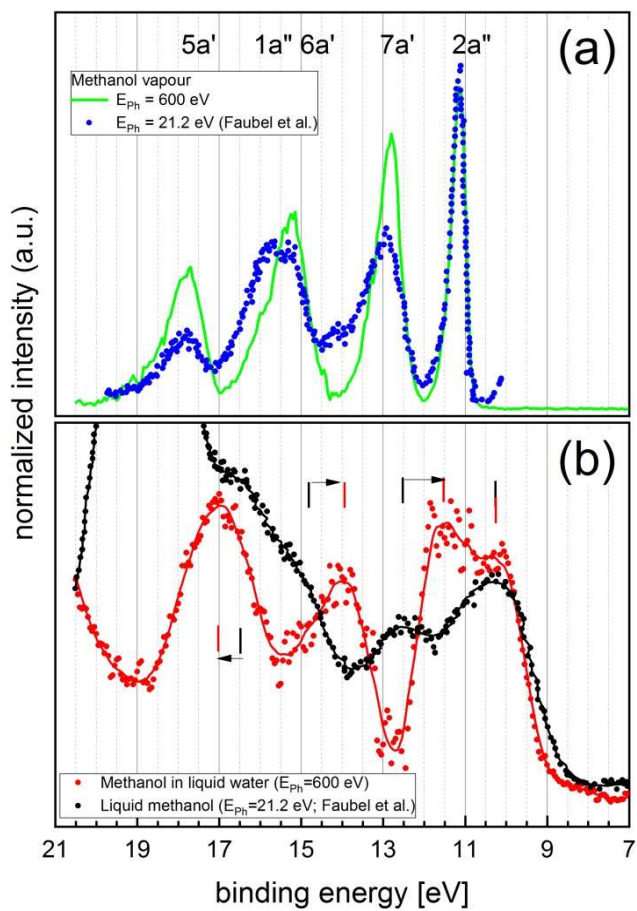

**Figure 7** Valence band XPS spectra of (a) methanol vapour (continuous green line) and (b) methanol in water (continuous red line and full red dots), respectively, taken from Fig. 6 (measured at 600eV photon energy). The spectra from pure methanol vapour (solid blue dots in a) and liquid (solid black dots and continuous black line in b) were taken from Faubel et al.[5] (measured at 21.22 eV photon energy) and have been rigidly shifted to match the 2a'' BE reported in this work. The short red and black vertical lines and arrows in (b) indicate the relative shifts and signs of the different MOs involved between these two spectra.

On the other hand, the gas methanol versus liquid methanol in water BE shifts from our present study shown in Table 6 are apparently inconsistent with the methanol gas-liquid BE shifts reported by Faubel et al. [5], except for the specific case of the low-lying 2a" peak. However, that former discrepancy does not come as a surprise as the two liquid phases are chemically distinctly different from each other; the resulting implications will be discussed below.

From the weighted subtraction of the water vapour VB data [see Fig. 3(a)] from the total *off-jet* VB raw data for obtaining the methanol vapour VB in Fig. 6(a), we can derive the intensity ratio between these two spectral constituents (see Fig. 9 in Annex A2). Based on this, we can calculate the vapour methanol volume fraction taking into account the average O2p orbital character of the methanol VB molecular orbitals of 46.6%. The vapour methanol volume fraction results as 81.1% (equivalent to a mole concentration $\chi_{vapour} = 0.48$), which is in good agreement with the analogue numbers derived from the *off-jet* O1s XPS data in the previous section (i.e., 82.5% methanol volume concentrations equivalent to $\chi_{vapour} = 0.50$).

In an analogue manner, the decomposition of the *on-jet* total XPS VB spectrum (see Fig. 10 in Annex A3) yields a methanol liquid volume fraction within the liquid µ-jet of 42.8%, equivalent to a methanol mole concentration $\chi_{liquid} = 0.25$. While the latter compares fairly well with the pertinent number for the sample liquid of $\chi_{liquid} = 0.21$, one has to take into account the finite XPS probing depth within the 30 µm cross section of the liquid µ-jet. Thus, taking into account the 600 eV photon energy, a photoelectron kinetic energy of about 580 eV in the present case of the VB measurements the inelastic mean free path (IMFP) for water is of the order of 17 Å. This indicates that for the present case of the methanol/water liquid mixture, the first outer layers of the liquid µ-jet exhibit an increase of the methanol volume fraction of up to roughly 43% as compared to the nominal bulk liquid value of 37.5%. This is in qualitative accordance to the results by Starr et al.[34] who conclude on a 43.5% methanol volume fraction ($\chi_{liquid} = 0.26$) for the same IMFP and sample solution based on *core level* XPS data from the liquid phase taken at various photon energies.

We now focus on the comparison between the XPS VB data from liquid methanol in water shown in Fig. 7(b) (red curve) taken at 600 eV as compared to the UPS VB data from liquid methanol (taken at HeI photon energy, black curve). What can be observed from the up to four peaks is:

(i)  BE shifts are different for the different MOs involved

(ii) A change in relative peak intensities with the same characteristics as could already be observed in the case of the methanol vapour VB spectra [Fig. 7(a)], mostly affecting the three lowest BE peaks.

Obviously, the second observation can readily be explained by the different PES cross sections the same way as in the case of the methanol vapour (see Figs. 13 and 14 in Annex B3). However, the observed BE shifts are more intricate as they are different for each peak/MO (see Table 7).

**Table 7** Experimental BE shifts from the comparison of the XPS VB spectra of liquid methanol as compared to methanol (37.5%)/water mixture shown in Fig. 7.

| VB peak | 2a" | 7a' | 6a' & 1a" | 5a' |
|---|---|---|---|---|
|  | ($n_0$) | ($\underline{n}_0$) | ($\sigma_{CO}$ & $\pi_{CH3}$) | ($\sigma_{OH}$) |
| BE shift experiment | 0.0 | −1.08 | −0.89 | 0.44 |

BE shift values are given in eV. Experimental BEs for liquid methanol are taken from [5] and have been shifted to match the 2a'' BE reported in this work.

In a first approach, these shifts can be attributed to a hydration of the methanol molecules within the methanol/water liquid solution. However, as a liquefaction process typically leads to an overall BE reduction due to, e.g., a hydration or a specific ordering of the molecules within the liquid phase we are apparently confronted with a more complex situation that affects different methanol molecular subgroups in different ways. Interestingly, it is the 5a' ($\sigma_{OH}$) peak related to the hydroxyl group that shows the deviating BE shift towards significant positive values, which is reminiscent of the earlier assumption by Faubel [5] that the hydroxyl group could be less affected than the methyl group in case the former would be oriented radially towards the outside of the liquid jet and thus exhibit a lesser influence by the liquid dielectric onto its BE – which would be consistent with our observation. However, due to the amphiphilic character of methanol, rather the methyl groups should point radially outside of the water/methanol mixture. This has been experimentally proven for liquid methanol/water mixtures [16] and clusters, where the results indicate the formation of a double layer including a "hydrophobic packaging" with an antiparallel oriented methanol molecules linked via an intermediate O-H chain encompassing water molecules

On the other hand, considering the methanol dimer formation proposed by Tomoda et al.[13] as a first initial step towards the liquid formation and comparing the VB spectral features in this study with the spectra shown in the lower panel of Fig. 7 we do not see any evidence for this from our experimental data.

### 3.2.3. DFT results for small $CH_3OH$ clusters

The O1s BE in single molecule methanol as predicted by DFT (see Table 2) is indeed somewhat lower (*i.e.*, by 0.7 eV) than for a single $H_2O$ molecule, which is in fair agreement with the corresponding experimental number of about 0.84 eV for the corresponding vapours (see Table 4). As in the case of water, methanol clustering decreases the O1s BE, which is associated with a resulting increase in O-H bond length. No clustering-induced shift in the C1s BE (of about 292 eV at DFT level – not shown) was predicted for the methanol models presented here. However, the latter can be due to the too small methanol cluster size, which probably does not correctly describe the weak, but apparently non negligible $CH_3$ group solvation, but rather only the OH hydrogen bonding within the cluster (see Fig. 12 in Annex B2).

Regarding the valence band PES region for the $CH_3OH \cdot (H_2O)_3$ cluster, the first peak (HOMO at -7.95 eV; see Fig. 12 in Annex B2; lower row, left column) is mainly due to the O2p lone pair, which may participate in electrostatic component of the hydrogen bond with neighbouring $CH_3OH/H_2O$ molecules. The experimental shifts to lower and higher BEs for the second (7a') and fourth (5a') peaks in the CH$_3$OH-H$_2$O mixture, respectively, as compared to pure liquid CH$_3$OH (see Fig. 7) can be tentatively explained due to their involvement in the covalent component of intermolecular hydrogen bonds (see Fig. 12, lower row): in pure methanol these states interact with each other, whereas in methanol-water mixture the CH$_3$OH-based 7a' states may interact with the H$_2$O-based 1b$_1$ states, which do have higher orbital energies (*i. e.,* lower BE) of -8.89 eV (H$_2$O 1b$_1$) and as compared to -9.53 eV (CH$_3$OH 7a'). In the same vein, the methanol 5a' states may mix with H$_2$O 1b$_2$ states at lower orbital energies (*i. e.,* higher BE) of -14.88 eV (H$_2$O 1b$_2$) as compared to -14.31 eV (CH$_3$OH 5a'). This corresponds fairly well with the experimental findings in Table 7. On the other hand, no BE shift was found in our DFT models for the third peak (i.e., 6a' and 1a" orbitals) in contrast to the experimental results. However, as the corresponding states are mainly localized at C-H and C-O bonds, this can be again assigned to the same issue as with C1s BE, namely the lack of proper simulation of CH$_3$ solvation effects in the present small cluster models.

## 4. Conclusions

The results from the present study shed some new light onto the electronic structure of water and water/methanol solutions. In more detail:

The results obtained from pure water at 600 eV photon energy emphasize the short range tetrahedral distribution already (pre-) existing in water and previously observed for crystalline and amorphous ice. We also find indications for ordering phenomena in water/methanol mixtures by the reduced O1s XPS liquid line width (i.e., as compared to pure water), very probably related to the amphiphilic character of the methanol molecule and which has already been apparent from earlier studies.

Regarding the C1s XPS lines from the water/methanol mixtures, we derive a gas/liquid BE shift of 0.95 eV while the vapour/liquid peak ratios allow for a quantitative determination of the methanol volume concentrations in both the vapour as well as in the liquid phase, that are corroborated by an analogue analysis of the valence band spectra.

A detailed quantitative analysis of the water/methanol liquid VB XPS spectrum accounting for the photon energy dependence of PES cross sections did confirm the atomic/orbital characteristics of the methanol molecular orbitals involved in the transitions and their pertinent intensities. From the decomposition of the liquid VB spectrum of the water/methanol mixture together with finite XPS probing depth we derive a methanol volume fraction of 43% for the outer liquid layers as compared to the nominal bulk liquid value of 37.5%.

Last but not least, from the different BE shifts of the water/methanol liquid VB spectrum with respect to that of pure methanol, we develop a CH$_3$OH•(H$_2$O)$_3$ cluster-based model that relates these different

BE shifts to the different MO hybridizations within that cluster, thus providing a qualitative explanation to the pertinent experimental valence band BE findings.

**Acknowledgements**    These experiments were performed at CIRCE beamline of the ALBA Synchrotron Light Facility with the collaboration of ALBA staff. The ICN2 is funded by the CERCA programme / Generalitat de Catalunya The ICN2 is supported by the Severo Ochoa Centres of Excellence programme, funded by the Spanish Research Agency (AEI, grant no. SEV-2017-0706) .This research was supported in part by the PLGrid infrastructure. Expert assistance by the Microliquids GmbH staff is gratefully acknowledged.

**Appendix A. Component analysis of XPS valence band spectra**

**A1. Component analysis of the water vapour/liquid *on-jet* XPS VB spectrum**

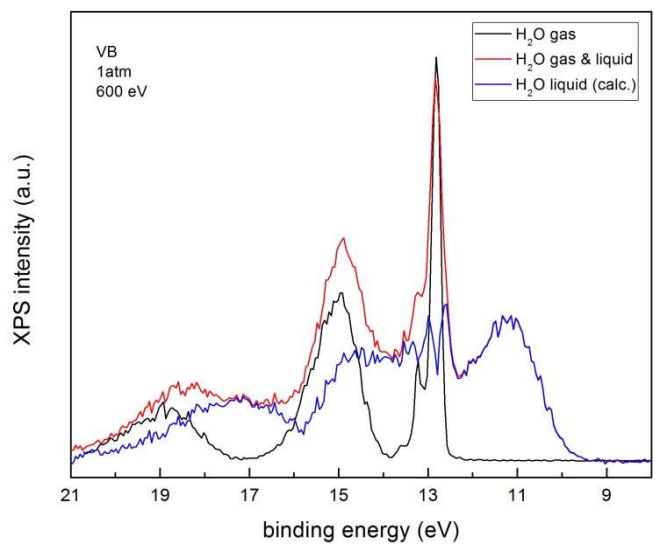

**Figure 8** Combined XPS valence band spectrum of water vapour and liquid water (continuous red line), water vapour (continuous black line), and the resulting difference spectrum for liquid water (continuous blue line). All data were taken at 600 eV photon energy.

**A2. Component analysis of the methanol/water vapour *off-jet* XPS VB spectrum**

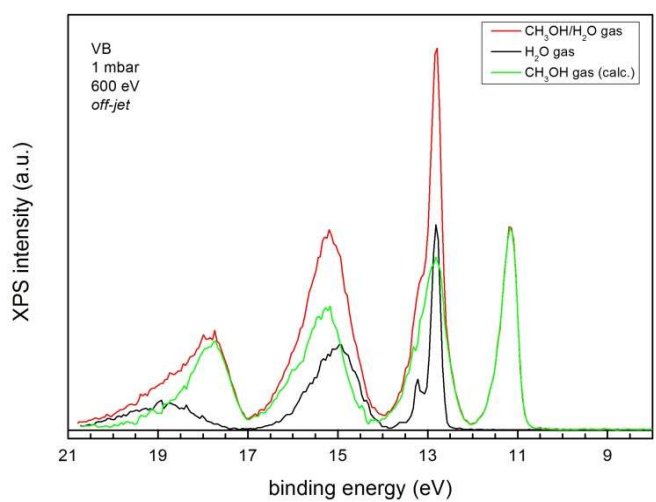

**Figure 9** Combined XPS valence band spectrum of water/methanol vapour (continuous red line), water vapour (continuous black line), and the resulting difference spectrum for methanol vapour (continuous green line). All data were taken at 600 eV photon energy.

## A3. Component analysis of the methanol/water *on-jet* XPS VB spectrum

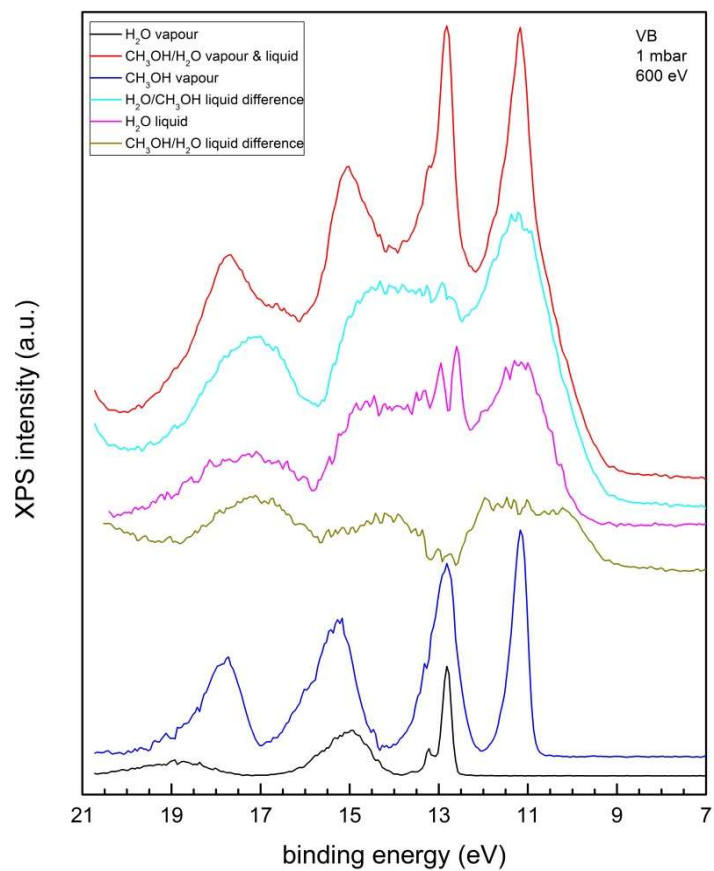

**Figure 10**   Combined XPS valence band spectrum of water/methanol vapour and liquid (continuous red line), water vapour (continuous black line), methanol vapour (continuous dark blue line), and the resulting difference spectrum for water/methanol liquid (continuous light blue line). XPS valence band spectrum of liquid water (continuous magenta line) and the resulting difference spectrum of liquid methanol in water (continuous ochre line). All data were taken at 600 eV photon energy.

**Appendix B. DFT molecular orbital calculations and XPS VB spectra**

**B1. Water molecular orbital calculations**

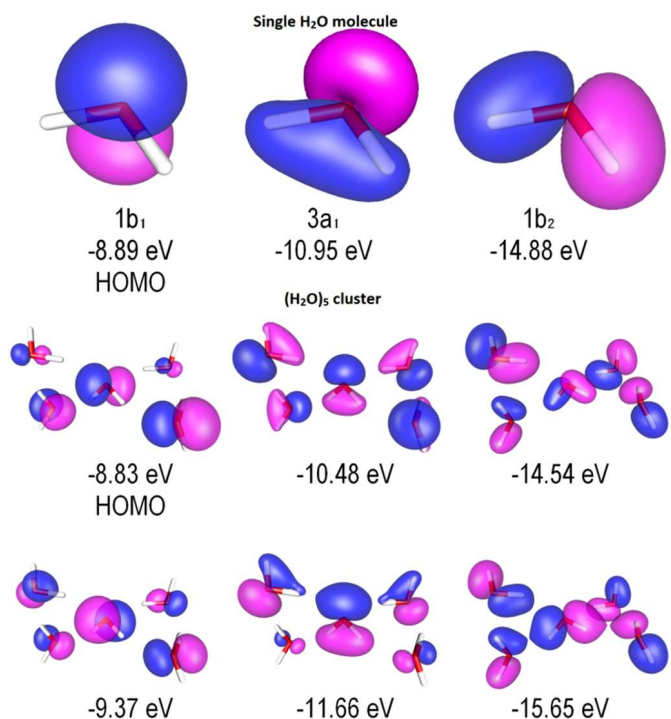

**Figure 11**  DFT calculated valence molecular orbitals and their orbital energies for a single water molecule (top panel) and for a $(H_2O)_5$ cluster (antibonding and bonding MOs in the centre and lower panel, respectively). For the $(H_2O)_5$ cluster, molecular orbitals largely localized at the central water molecule in tetrahedral coordination are depicted.

**Table 8**  Numerical results from DFT calculations on single water molecule

| Molecular Orbital // Orbital energy (eV), Orbital description | Atomic orbital contributions to MO |
|---|---|
| $1a_1$ // -522.50 (O1s core level) | **O1s 100%** |
| $2a_1$ // -28.27 (O2s core level) | **O2s 90%,** H1s 6%, O2p 4% |
| $1b_2$ // -14.88 (O-H bond) | **O2p 68%, H1s 32%** |
| $3a_1$ // -10.95 (O lone pair + O-H bond) | **O2p 67%, O2s 23%,** H1s 9% |
| $1b_1$ // -8.89 (O lone pair, HOMO) | **O2p 100%** |

## B2. Methanol molecular orbital calculations

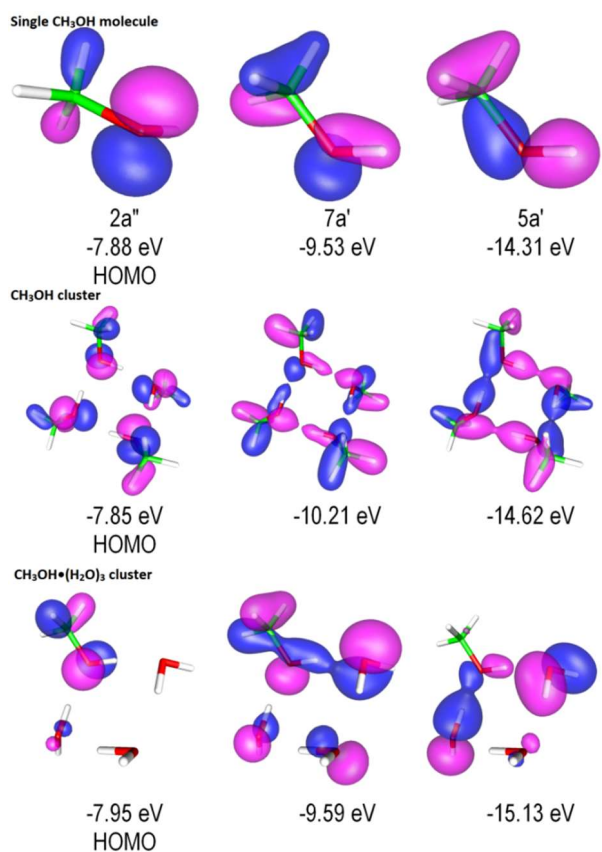

**Figure 12**    DFT calculated molecular orbitals and their orbital energies for $CH_3OH$, which participate in hydrogen bonds in the condensed phase: Single methanol molecule (top panel), methanol tetramer (centre panel) and $CH_3OH•(H_2O)_3$ cluster (bottom panel).

**Table 9**   Numerical results from DFT calculations on single methanol molecule

| Molecular Orbital//Orbital energy (eV), Orbital description | Atomic orbital contributions to MO |
|---|---|
| 1a' // -522.53 (O1s core level) | **O1s 100%** |
| 2a' // -279.16 (C1s core level) | **C1s 100%** |
| 3a' // -28.58 (O2s core level) | **O2s 83%**, C2s 7%, H1s 3% |
| 4a' // -19.12 (C2s core level) | **C2s 61%**, H1s 15%, O2p 11%, O2s 10%, |
| 5a' // -14.31 (C-O and O-H bond) | **O2p 47%, H1s 21%, C2p 20%**, O2s 6% |
| 1a" // -12.36 (C-H bond) | **C2p 46%, H1s 28%, O2p 26%** |
| 6a' // -12.01 (C-H and C-O bond) | **O2p 40%, C2p 38%**, H1s 16%, O2s 3% |
| 7a' // -9.53 (mainly O-H bond) | **O2p 44%, H1s 24%**, C2p 16%, O2s 7% |

| 2a' // -7.88 (O2p lone pair, HOMO) | **O2p 76%**, H1s 17%, C2p 6% |

**B3. Methanol molecule VB spectra as a function of photon energies and photoemission cross sections**

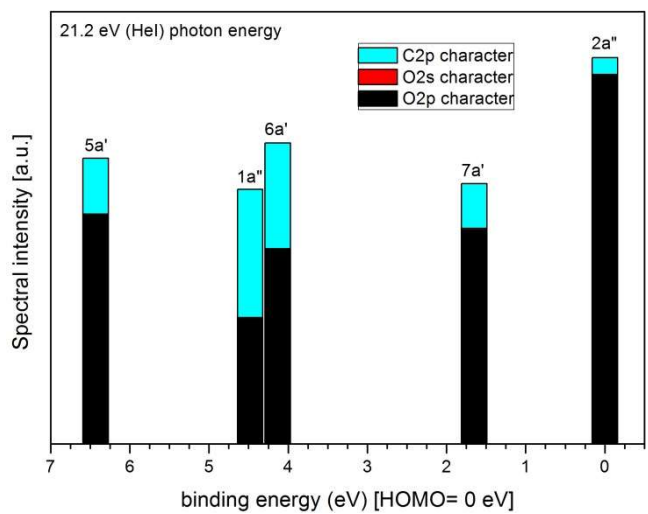

**Figure 13**    Calculated methanol molecule VB spectrum at HeI photon energy (21.22 eV)

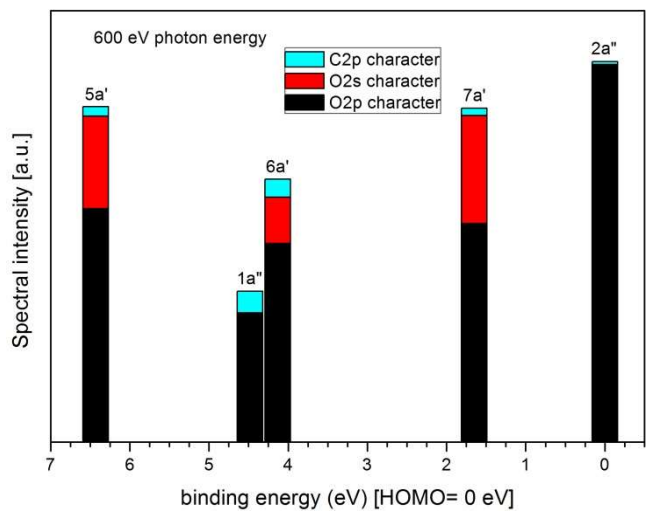

**Figure 14**    Calculated methanol molecule VB spectrum at 600 eV photon energy